\documentclass[twocolumn,showpacs,preprintnumbers,amsmath,amssymb]{revtex4}

\usepackage{graphicx}
\usepackage{dcolumn}
\usepackage{bm}

\begin{document}

\preprint{APS/123-QED}

\title{Four-fold structure of vortex core states in Bi$_2$Sr$_2$CaCu$_2$O$_{8+\delta}$}

\author{Giorgio Levy}
 \email{giorgio.levy@physics.unige.ch}
\author{Martin Kugler}
\author{Alfred A. Manuel}
\author{$\O$ystein Fischer}
\affiliation{DPMC, Universit\'e de Gen\`eve, 24 Quai
Ernest-Ansermet, CH-1211 Gen\`eve 4, Switzerland}
\author{Ming Li}
\affiliation{Kamerlingh Onnes Laboratory, Leiden University, P.O.
Box 9506, 2300 RA Leiden, The Netherlands}

\date{\today}

\begin{abstract}
We present a detailed study of vortex core spectroscopy in
slightly overdoped Bi$_2$Sr$_2$CaCu$_2$O$_{8+\delta}$ using a low
temperature scanning tunneling microscope. Inside the vortex core
we observe a four-fold symmetric modulation of the local density
of states with an energy-independent period of ($4.3\pm0.3)a_0$.
Furthermore we demonstrate that this square modulation is related
to the vortex core states which are located at $\pm6$~meV. Since
the core-state energy is proportional to the superconducting gap
magnitude $\Delta_p$, our results strongly suggest the existence
of a direct relation between the superconducting state and the
local electronic modulations in the vortex core.
\end{abstract}


\maketitle

The high-temperature superconductors are characterized by an
unconventional temperature-doping phase diagram, which is the
object of numerous studies focusing in particular on the nature of
the pseudogap and superconducting states. However, in spite of
this effort the microscopic origin of superconductivity in these
materials is still not understood. A promising approach to
investigate the superconducting state is to study the electronic
properties of the vortex cores. Scanning tunneling microscopy
(STM) observations of vortex cores were first carried out on
NbSe$_2$ \cite{Hess-1989}. The behavior of the tunneling
conductance, which measures the local density of states (LDOS),
was found to agree with the prediction by Caroli {\it et al.}
\cite{Caroli-1964}, that a band of localized states develops in
the cores. The subsequent observation of vortices in
YBa$_2$Cu$_3$O$_{7-\delta}$ (YBCO) \cite{Maggio-Aprile-1995} gave
a surprising result: contrary to the observations in NbSe$_2$, the
vortex-core spectra showed two peaks staying at a constant energy
irrespective of the distance to the vortex center, as if the
vortex would contain only two localized states instead of a whole
band. Following this, several groups investigated theoretically
the vortex core in a $d$-wave superconductor, leading to the
conclusion that the spectra observed in YBCO cannot be explained
in the framework of the BCS theory \cite{Wang-1995} and that an
extension of this theory is necessary at the very least
\cite{Franz-1999_Berthod-2001b_Zhu-2001a}. The STM study of
Bi$_2$Sr$_2$CaCu$_2$O$_{8+\delta}$ (Bi2212) uncovered another
property of the vortex-core spectra: they display the low
temperature pseudogap \cite{Renner-1998b}. Subsequently it was
found that the two localized states are also present in this
compound \cite{Pan-2000b_Hoogenboom-2000a_Matsuba-2003}, and that
the states seen in YBCO and in Bi2212 have a common
characteristic, appearing at an energy of about $0.3\Delta_p$,
where $\Delta_p$ is the position of the coherence peaks in the
superconducting state \cite{Hoogenboom-2001a}.

\begin{figure*}[htp]
\centering
\includegraphics[width=18cm,clip]{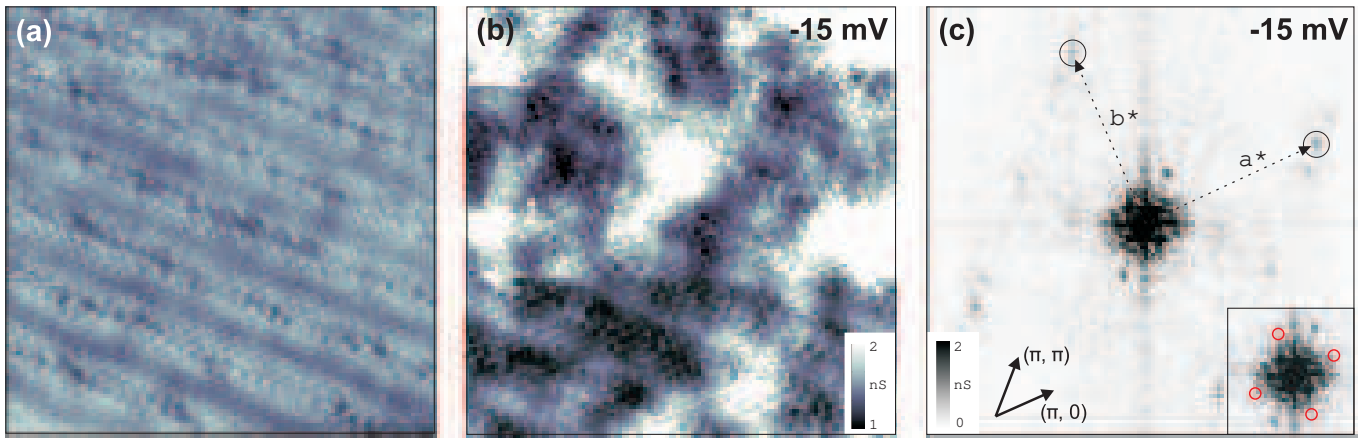}
\caption{\label{fig:ZeroField} (a) $16 \times 16$~nm$^2$
topographic image of Bi2212 showing the atomic lattice on the BiO
plane. The atomic corrugation is $2$~\AA. The image was acquired
at a bias voltage $V=0.6$~V and a tunnel current $I=0.8$~nA. (b)
$dI/dV$ conductance map in the same area at $V=-15$~mV, zero field
and $T=2$~K. (c) Absolute value of the Fourier transform of (b).
$\bm{a}^*$ and $\bm{b}^*$ are the reciprocal vectors of the atomic
lattice with modulus $2\pi/a_0$ and $2\pi/b_0$, respectively
($a_0\simeq b_0=3.8$~\AA). The position where peaks corresponding
to a $4a_0 \times 4a_0$ period would be expected are indicated in
red in the inset.}
\end{figure*}

Hoffman {\it et al.} observed a spatial modulation of the low
energy tunneling conductance in the vortex cores having a period
of about $4a_0$ \cite{Hoffman-2002a}. Later, it was found that
similar modulations also appear in the absence of magnetic field
\cite{Hoffman-2002b,Howald-2003}. Hoffman {\it et al.} noticed
that the wavelength of these modulations disperse with energy and
it was proposed that the effect can be understood in terms of
quasiparticle interference due to scattering on impurities and
other inhomogeneities \cite{Hoffman-2002b,Wang-2003}. However,
some of the periodic modulations reported by Howald {\it et al.}
\cite{Howald-2003} did not disperse in energy, and an explanation
in terms of static stripes was put forward. More recently,
Vershinin {\it et al.} \cite{Vershinin-2004} studied the spatial
dependence of the tunneling conductance in the pseudogap phase.
They observed an incommensurate square lattice with period
$(4.7\pm0.2)a_0$. The modulations observed above $T_c$ do not
disperse and it was thus concluded that they are different from
the interference modulations seen in the superconducting state. A
non-dispersing square pattern was also reported at low
temperatures in strongly underdoped Bi2212 \cite{McElroy-2004a}
and Ca$_{2-x}$Na$_x$CuO$_2$Cl$_2$ (NCCOC) \cite{Hanaguri-2004},
which are characterized by pseudogap-like spectra. Thus there is
considerable evidence that a square pattern with a non-dispersing
wavelength is associated with the pseudogap, whereas in the
superconducting state one observes predominantly dispersing
modulations, presumably due to quasiparticle interference.

In this letter we report a detailed study of the LDOS modulation
inside the vortex core. We confirm the early observations by
Hoffman {\it et al.} \cite{Hoffman-2002a}, but our measurements
show in addition that this modulation does not disperse with
energy, like the ones observed in the pseudogap phase
\cite{Vershinin-2004}. We further demonstrate that this square
modulation is linked to the localized vortex-core state
\cite{Maggio-Aprile-1995,Pan-2000b_Hoogenboom-2000a_Matsuba-2003},
and we thereby establish a direct relation between the vortex core
electronic modulations and the superconducting state.

Our STM measurements were performed on a Bi2212 single crystal
grown by the travelling solvent flux zone method and annealed in
$500^{\circ}$C under $15$~bar oxygen pressure. After annealing we
measured $T_c^{onset}=88$~K ($\Delta T_c=4$ K) by
ac-susceptibility. The relatively flat background slope of the
conductance spectra, as well as the magnitude of the
superconducting gap $\Delta_p$ indicate that the sample is
slightly overdoped. In the region studied, we observe an average
gap $\bar\Delta_p=25.2$~meV with a standard deviation
$\sigma=4$~meV. This gap distribution is consistent with the
observed superconducting transition width \cite{Hoogenboom-2003a}.

We performed the measurements with a home-built STM
\cite{Kugler-2000a} under ultrahigh vacuum. The sample was cleaved
{\it in situ} at room temperature before cooling and applying the
magnetic field and the tunnel junction was made between the (001)
sample surface and an electrochemically etched Iridium tip. All
data presented here was acquired at 2~K, first at zero field and
then at 6~T.

In Fig.~\ref{fig:ZeroField}a we show a topographic scan at zero
field which clearly resolves the atomic lattice of the BiO top
layer, as well as the characteristic supermodulation running along
the (1,1) direction. Figure~\ref{fig:ZeroField}b is a conductance
map at $V=-15$~mV, which was acquired in the same area
simultaneously with the topographic image. The Fourier transform
(FT) shown in Fig.~\ref{fig:ZeroField}c reveals four peaks
corresponding to the Bi lattice and several peaks due to the
supermodulation and its harmonics running along the ($\pi$,$\pi$)
direction. In contrast to earlier reports
\cite{Hoffman-2002b,Howald-2003,Vershinin-2004} we do not observe
quasiparticle interference peaks in zero field, presumably because
of the absence of sufficiently strong scattering centers. The red
circles in the inset of Fig.~\ref{fig:ZeroField}c indicate the
positions where the peaks of a $4a_0 \times 4a_0$ modulation would
have been located.

\begin{figure}[htp]
\centering
\includegraphics[width=8.2cm,clip]{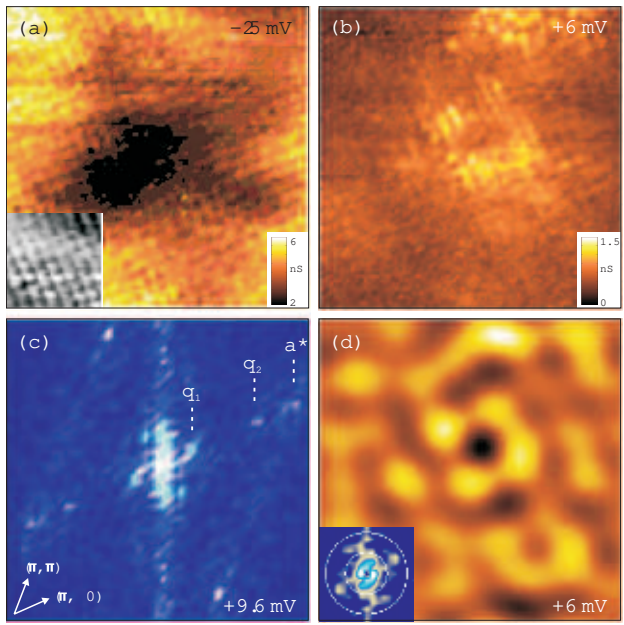}
\caption{\label{fig:Core} (a) $8.7 \times 8.7$ nm$^2$ conductance
map at $V=-25$~mV. The inset shows the simultaneously acquired
topography at the same scale as the underlying conductance map.
(b) Conductance map in the same area as (a) at $V=+6$~mV. (c) FT
image at $V=+9.6$~mV. (d) Filtered inverse FT image. The inset
shows the filter applied to the image acquired at $V=+6$~mV and
which selects the region between the two circles with radii
$q[2\pi/a_0]\sim0.17$ and $\sim0.32$.}
\end{figure}

\begin{figure}[h!]
\centering
\includegraphics[width=8.2cm,clip]{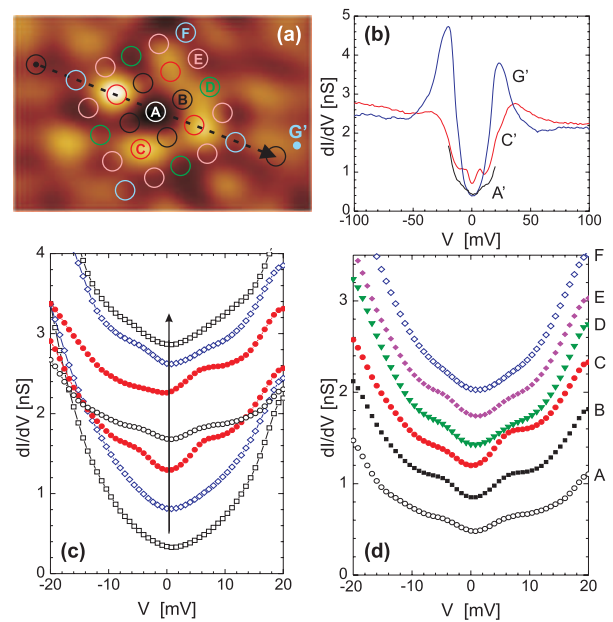}%
\caption{\label{fig:Analysis} (a) Central region of Fig.
\ref{fig:Core}d. Each circle contains 21 pixels. (b) Spectra taken
at the core center (A'), on a maximum of the square pattern (C')
and outside the core (G'). (c) Spectra averaged in the 7 circles
along the arrow in (a); For clarity, the spectra were shifted
vertically by $0.4$~nS. (d) Spectra averaged over the four-fold
symmetry equivalent circles, drawn with identical color in (a);
The spectra are offset vertically by $0.3$~nS.}
\end{figure}

We now turn to the measurement performed at 6~T and focus on a
vortex core. As shown in Fig.~\ref{fig:Analysis}b, the vortex core
spectra are characterized by a gap-like structure similar to the
pseudogap measured above $T_c$ \cite{Renner-1998b} and weak
low-energy structures attributed to localized quasiparticle
excitations \cite{Pan-2000b_Hoogenboom-2000a_Matsuba-2003}. In
Fig.~\ref{fig:Core}a we show a conductance map at $V=-25$~mV which
clearly displays the location of the vortex core. Its size and
irregular shape are consistent with previous studies
\cite{Renner-1998b,Hoogenboom-2000b}. In Fig.~\ref{fig:Core}b we
present a conductance map at $V=+6$~mV which corresponds to the
energy of the core state. Inside the core one can observe a
striking square pattern formed by four bright regions, similar to
the observations by Hoffman {\it et al.} \cite{Hoffman-2002a}.

In order to quantitatively analyze these structures we performed
the FT of conductance maps at several energies. In
Fig.~\ref{fig:Core}c we show the FT obtained at $V=+9.6$~mV. In
addition to the peaks corresponding to the atomic lattice, we
observe two clear structures. First we see four peaks at
$q_1\simeq 0.25 \pi/a_0$ corresponding to an incommensurate period
of $(4.3\pm 0.3)a_0$ oriented parallel to the CuO bond direction.
These maxima are clearly visible in all LDOS-FT taken  between $4$
and $12$~mV, and between $-8$ and $-12$~mV. We note that their
intensity at negative bias (occupied states) is $\sim 2/3$ smaller
than at positive bias (empty states). Second we see two maxima at
$q_2\simeq 0.75 \pi/a_0$ which only appear along the $(\pm\pi,0)$
direction. Looking closer, one can observe that the quartet of
$q_1$ peaks is slightly rotated with respect to the atomic
lattice, while the two $q_2$ peaks are not.

To clearly identify which signal in the real-space conductance map
of Fig.~\ref{fig:Core}b originates from the $q_1$ peaks, we show
in Fig.~\ref{fig:Core}d the filtered inverse FT. We selected a
region in $q$-space containing the four $q_1$ peaks (see inset).
The inverse FT exhibits four bright regions at the corners of a
square which clearly correspond to the pattern observed in the raw
data. We thus demonstrate that the low energy structure in the
vortex core shown in Fig.~\ref{fig:Core}b is indeed at the origin
of the four $q_1$ peaks.

We now address the spatial variation of the LDOS inside the vortex
core. In Fig.~\ref{fig:Analysis}a we indicate 27 circular areas,
each of containing 21 pixels of our spectroscopic image. For
clarity we used the filtered image (Fig.~\ref{fig:Core}d) to
identify the positions. In Fig.~\ref{fig:Analysis}b we display
three spectra: A' and C' were taken at the center of area A and C,
respectively and G' was taken at point G, at a distance of
$3.4$~nm from the center of the square pattern. Whereas G' is
similar to the zero field spectrum, A' and C' differ in a
remarkable manner. The core states appear very distinctly in
spectrum C', which is at one maximum of the fourfold pattern, but
in A' there is hardly any signature of the core states. Thus it
appears that the square pattern reflects the spatial variation of
the localized states. In order to investigate this further and
extract the most robust features we have averaged spectra inside
each circle of Fig.~\ref{fig:Analysis}a.
Figure~\ref{fig:Analysis}c shows the average spectra of each
circle along the trace indicated by the arrow in
Fig.~\ref{fig:Analysis}a. The core states appear clearly in the
two circles located on the maxima of the square pattern (red). At
the center A, only a weak signature of the localized states is
seen, and this signature disappears when moving outside the
fourfold pattern. In Fig.~\ref{fig:Analysis}d we show a different
representation of this data. Whereas A again shows the average
spectrum of the central circle, the curves B--F correspond to an
average of spectra taken in circles which are equivalent by the
four-fold symmetry (same color in Fig.~\ref{fig:Analysis}a) and at
increasing distance from the center. The strongest signature of
core states is again seen in curve C taken on the maxima of the
square pattern.

\begin{figure}[htp]
\centering
\includegraphics[width=7cm,clip]{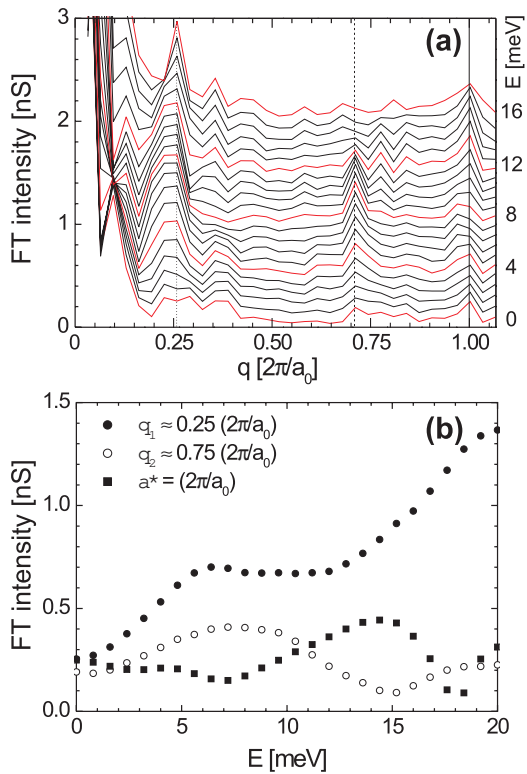}
\caption{\label{fig:Dispersion} (a) Tunneling conductance at
energies between 0 and $+16$~meV measured along $(\pi,0)$. (b) FT
intensity of $q_0=2\pi/a_0$, corresponding to the atomic lattice,
$q_1\simeq0.25(2\pi/a_0)$ corresponding to the square pattern in
the vortex core, and $q_2\simeq0.75(2\pi/a_0)$ which only appears
along the $(\pm\pi,0)$ directions. The intensities where measured
on the peaks indicated in Fig.~\ref{fig:Core}c.}
\end{figure}

We therefore conclude that the vortex-core states are closely
related to the square pattern. In Fig.~\ref{fig:Dispersion}a we
plot a cut in the Fourier transform along the $(\pi,0)$ direction
at several energies. The peak at $q_1$ corresponding to the
$4a_0$-period does not disperse with energy within the error
margins of our measurement. While a weak dispersion towards longer
wavelengths cannot be excluded given our present resolution, this
possible dispersion would be in the opposite sense than the one
observed along $(\pi,0)$ in the superconducting state and
attributed to quasiparticle interference \cite{Hoffman-2002b}. We
also remind that the peak at $q_2$ corresponding to a wavelength
of $(4/3)a_0$ is present in the $(\pi,0)$ direction, but not in
the $(0,\pi)$ direction.

Figure~\ref{fig:Dispersion}b shows the intensities of the three
dominant peaks as a function of energy. The intensity of the $q_1$
peak has a clear maximum at the energy of the localized state and
increases again when approaching $20$~meV. In fact this curve
mimics the local density of states measured at the four maxima of
the square pattern (Fig.~\ref{fig:Analysis}d, curve C).

Comparing our results with the observations of Vershinin {\it et
al.} \cite{Vershinin-2004} in the pseudogap state above $T_c$, we
find that the ordering in the vortex core is very similar to the
ordering in the pseudogap state. The basic structure is a
non-dispersing square modulation in the Cu-O bond direction. The
period $(4.3\pm0.3)a_0$ in the vortex core at 2~K is slightly
smaller than the period $(4.7\pm 0.2)a_0$ in the pseudogap state
at 100~K. In the pseudogap state, the intensity of the peak in the
FT was found to be largest and energy independent below $20$~meV,
whereas we find an energy dependence resembling the tunnel
conductance. This discrepancy may be due to temperature broadening
at 100~K. Another difference is that we find a $(4/3)a_0$
modulation in one direction. Such ordering was also seen by
Hanaguri {\it et al.} \cite{Hanaguri-2004} at low temperature in
Ca$_{2-x}$Na$_x$CuO$_2$Cl$_2$ with a main period of $4a_0$.

In this paper we evidence that the amplitude of the vortex core
states has a four-fold structure directly reflecting the
modulation observed in the vortex core. Since these states appear
at an energy proportional to the gap $\Delta_p$, our results
connect the superconducting state to the electronic modulation. We
further find that the four-fold modulation has the same behaviour
as in the pseudogap phase \cite{Vershinin-2004}, what could be
expected since the vortex cores display the pseudogap
\cite{Renner-1998b}. The relation between the pseudogap and the
superconducting state has been the topic of many theoretical
studies possibly leading to spatially modulated structures
\cite{Zaanen-1989_Zhang-1997b_Emery-1999_Voijta-1999_Wen-1996_Chakravarty-2001}.
In particular, several authors have proposed that a pair density
wave (PDW) is at the origin of the observed structures
\cite{Chen-2002_Chen-2004a_Tesanovic-2004_Anderson-2004b}. In the
context of our study this model is attractive, since we establish
a clear link between the superconducting state, the pseudogap and
the square pattern in the vortex core. Within the PDW picture,
this suggests that the localized states at $E\approx0.3\Delta_p$
correspond to the lowest pair breaking excitations of the PDW. The
relation between the localized states and the square pattern thus
sets a critical test for these theories.

This work was supported by the Swiss National Science Foundation
through the National Centre of Competence in Research "Materials
with Novel Electronic Properties - MaNEP". We acknowledge
C.~Berthod for very useful discussions and pertinent suggestions
and A.~P.~Petrovic for carefully reading the manuscript.

\end{document}